\documentclass[sigplan,10pt]{acmart}
\renewcommand\footnotetextcopyrightpermission[1]{}
\usepackage{subcaption}
\AtBeginDocument{%
  }

\begin{document}

\title{Invisible Yet Dominant: Big Stalls of Kernel I/O Mechanisms in Cloud OLTP Databases}

\author{Mitsumasa Kondo}
\affiliation{%
  \institution{NTT, Inc.}
  \city{}
  \country{}
}
\email{mitsumasa.kondou@ntt.com}

\settopmatter{printfolios=false,printacmref=false}
\maketitle
\pagestyle{empty}

\section{Introduction}
The standard block device in the cloud is distributed block storage.
Because cloud providers must manage their infrastructure efficiently while guaranteeing data durability, direct-attached storage (DAS) offered by most cloud services is volatile: it cannot be used by applications such as databases that require persistence and durability.
Consequently, compute--storage disaggregated databases~\cite{verbitskiAmazon2017, antonopoulosSocrates2019, noauthorAlloydbNodate, depoutovitchTaurus2020, pangUnderstanding2024}---physically separating the compute layer from the storage layer---have become the dominant architecture for high-performance cloud databases.
The driving reason is a well-known phenomenon: when a database that issues many random writes runs on top of distributed block storage, its I/O path stalls and high performance cannot be achieved.
The core idea of disaggregation is to provision database-dedicated storage and to write only the database's Write-Ahead Log (WAL) directly to it, bypassing the kernel I/O layer; database pages are then reconstructed from the WAL on the storage side, eliminating random writes.
Despite the success of this approach, the random-write stalls themselves have remained poorly understood.
They have conventionally been attributed to network PPS ceilings, bandwidth limitations, and general kernel-layer overhead.

\begin{figure}[t]
\centering
\includegraphics[width=0.85\linewidth]{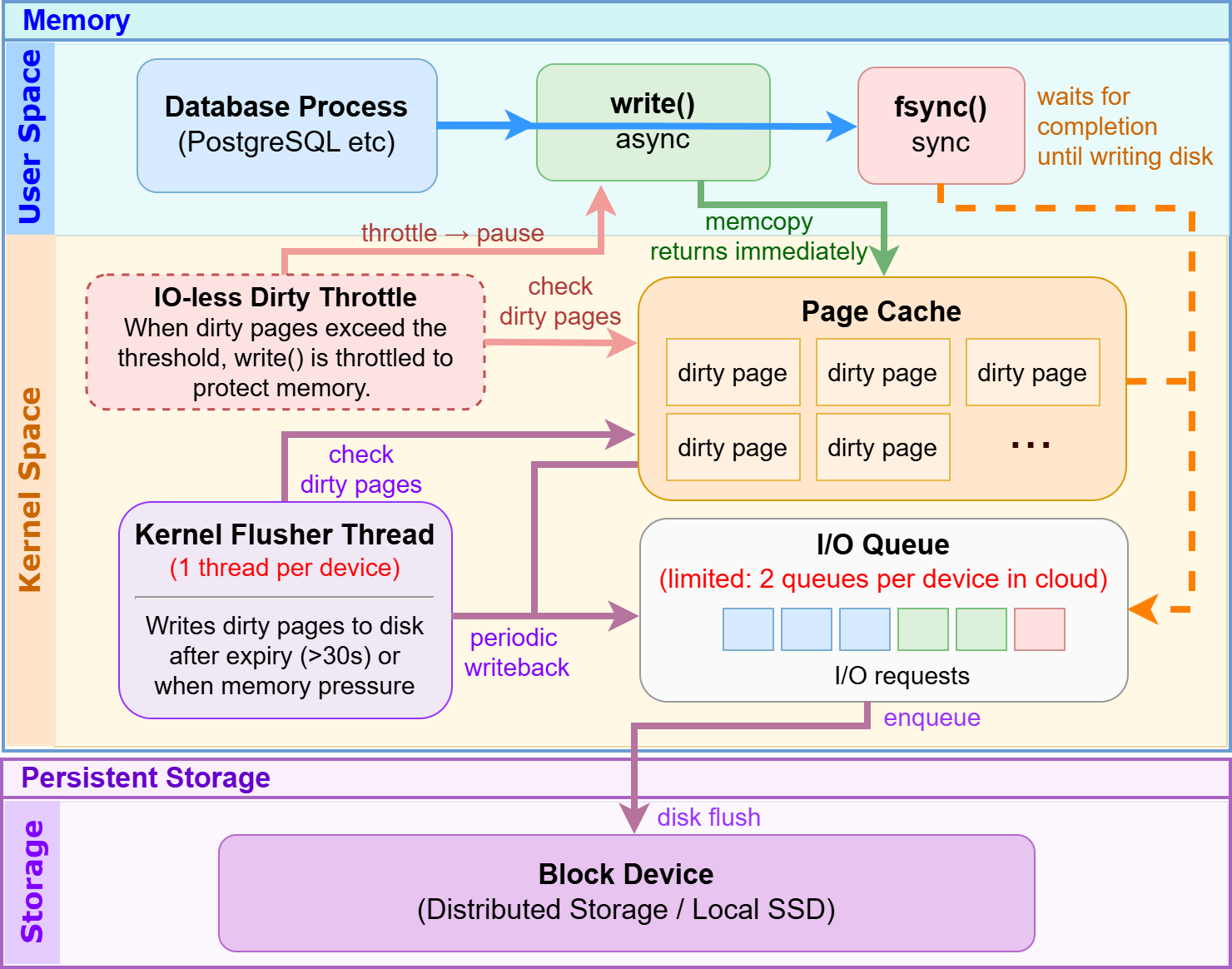}
\vspace{-9pt}
\caption{Linux Kernel Writeback Architecture} 
\vspace{-9pt}
\label{fig:linux-writeback}
\vspace{-0.3cm}
\end{figure}

SteelDB~\cite{kondo2026steeldbdiagnosingkernelspacebottlenecks} confronted this folklore head-on: why do random writes actually stall on distributed block storage?
It discovered that the root cause is a design misalignment across three layers---the distributed block storage, the kernel, and the database---and proposed the SteelDB architecture, a cross-layer orchestration zero-patch architecture that achieves high performance without modifying any of these layers.
Specifically, the kernel's flusher thread (KFT), responsible for writing dirty pages back to disk, operates as a single thread per block device.
On distributed block storage, whose per-I/O latency is inherently higher due to network-attached access and data replication, this single KFT cannot drain dirty pages fast enough, triggering the kernel's memory protection mechanism---IO-less Dirty Throttle~\cite{fengguangFengguangNodate}---which forcibly pauses foreground write system calls.
Furthermore, distributed block storage typically exposes only two I/O queues per device for multi-tenant QoS isolation, and this limited number of queues exacerbates congestion when the single KFT cannot drain dirty pages promptly.
SteelDB resolves these bottlenecks, achieving 3.1$\times$ the throughput of Amazon Aurora~\cite{verbitskiAmazon2017} and 1.4$\times$ that of Google AlloyDB~\cite{noauthorAlloydbNodate} on TPC-C at less than half the cloud cost.
Its advantage is not limited to runtime metrics: an analysis of historical release records shows that Aurora takes a median of 254 days to port its proprietary patches to a new PostgreSQL major version, versus zero days for the patch-free SteelDB.

Whereas SteelDB evaluated performance from the database perspective, this paper shifts focus to the kernel perspective.
We use eBPF to analyze, from inside the kernel, the bottlenecks that SteelDB resolved.
These stalls are largely invisible to standard counters such as \texttt{/proc/diskstats}, and we quantify their impact on cloud OLTP performance.

\begin{figure*}[t]
  \centering
  \begin{subfigure}{\textwidth}
    \centering
    \includegraphics[width=1.00\linewidth]{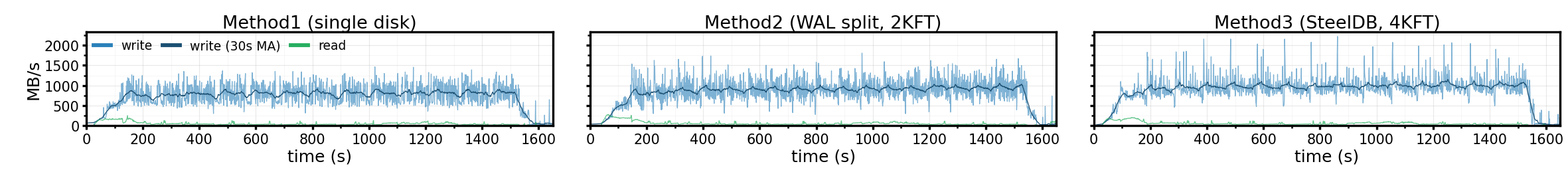}
    \vspace{-22pt}
\caption{Disk I/O bandwidth (total provisioned: 80K IOPS, 2.0\,GB/s) via \texttt{block:block\_rq\_issue/complete}.}
    \vspace{-2pt}
    \label{fig:iostat}
  \end{subfigure}
  \vspace{0.12em}
  \begin{subfigure}{\textwidth}
    \centering
    \includegraphics[width=1.0\linewidth]{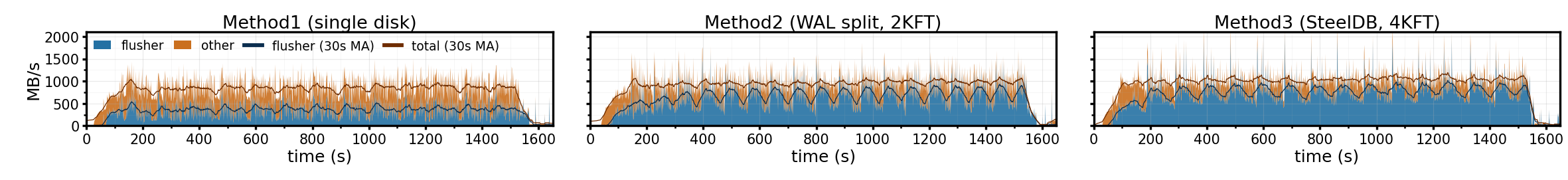}
    \vspace{-22pt}
\caption{Disk I/O bandwidth by issuing context (KFT writeback vs.\ foreground), via \texttt{writeback:writeback\_start/written}.}
    \vspace{-3pt}
    \label{fig:kft}
  \end{subfigure}
  \vspace{0.45em}
  \begin{subfigure}{\textwidth}
    \centering
    \includegraphics[width=1.0\linewidth]{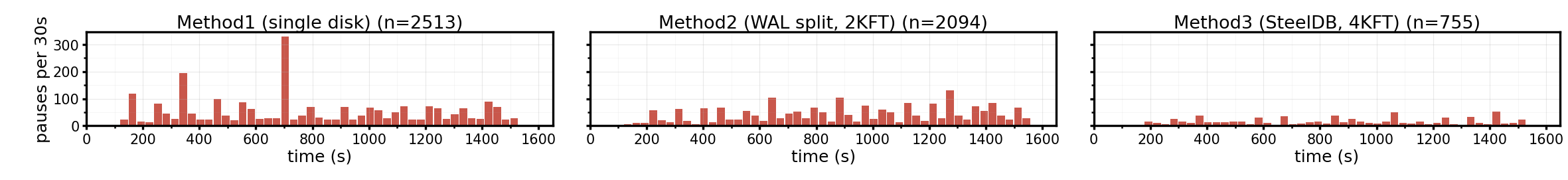}
    \vspace{-22pt}
    \caption{IO-less Dirty Throttle pause events captured via \texttt{writeback:balance\_dirty\_pages}.} 
    \vspace{-13pt}
    \label{fig:pause}
  \end{subfigure}
\end{figure*}

\vspace{-0.2cm}
\section{Analysis of Kernel I/O Bottlenecks}
\label{sec:analysis}
 
\noindent\textbf{SteelDB's cross-layer resolution.}\quad
Figure~\ref{fig:linux-writeback} illustrates the kernel I/O path from a database process to distributed block storage, including the bottlenecks described in section 1.
SteelDB resolves these by provisioning multiple disks and strategically placing database data across them, thereby multiplying both the number of KFTs and I/O queues while physically isolating their I/O paths.
This enables parallel dirty-page writeback, alleviates I/O queue congestion, suppresses IO-less Dirty Throttle, and delivers high database performance.
The SteelDB architecture is the product of a cross-layer exploration spanning the kernel's internal I/O characteristics, the architecture of distributed block storage, and the I/O patterns of the database.
For further details we refer the reader to the original paper.
 
\noindent\textbf{eBPF-based profiling.}\quad
We developed an eBPF-based tool that visualizes kernel I/O-path stalls invisible to standard counters.
It attaches to the \texttt{block:block\_rq\_issue} and \texttt{blo
ck:block\_rq\_complete} tracepoints to capture every request traversing the block layer, from which it monitors the aggregate disk I/O bandwidth over time (Figure~\ref{fig:iostat}).
To classify each request by issuing context---separating
KFT-originated writeback from other contexts such as foreground
fsync (Figure~\ref{fig:kft})---it tags each bio with its issuing
context at submission time. It then correlates these events with
the \texttt{writeback:writ
eback\_start}/\texttt{writeback\_written}
tracepoints, which mark the intervals during which a kworker is
executing \texttt{wb\_writeb
ack()}.
In addition, to capture IO-less Dirty Throttle pauses---intentional delays of write system calls when dirty pages exceed the threshold---we attach to the \texttt{writeback:balance\_di
rty\_pages} tracepoint and aggregate invocations into a time-series bin graph (Figure~\ref{fig:pause}).

\begin{table}[t]
  \centering
  \caption{TPC-C Benchmark Results and Disk I/O.}
  \vspace{-8pt}
  \label{tab:tpcc}

  \small
  \setlength{\tabcolsep}{3pt}
  \begin{tabular}{lrrrr}
    \toprule
    Method & Throughput & Ave Write. & Ave Read. & Max Trans\\
           & (NOPM) & (MB/s) & (MB/s) &  Lat. (ms) \\
    \midrule
    1: Single disk & 444,326 & 679.4 & 50.7 & 1,238.2 \\
    2: WAL split   & 511,480 & 754.1 & 62.1 &   848.1 \\
    3: SteelDB     & 544,354 & 827.1 & 60.2 &   502.5 \\
    \bottomrule
  \end{tabular}
  \vspace{-12pt}
\end{table}
 
\noindent\textbf{Evaluation setup.}\quad
All experiments were conducted on AWS using Rocky Linux~9.8 with Linux kernel~5.14.
The database was PostgreSQL~16.4, and TPC-C~\cite{noauthorTpcCNodate} benchmarks were driven by HammerDB~\cite{noauthorHammerdbNodate}.
The VM instance type was c6in.8xlarge (32~vCPUs, 64\,GB RAM).
The database size was 1K~WH (approximately 100\,GB) with 256~VU.
We compared three disk configurations with identical total provisioned IOPS (80K) and bandwidth (2.0\,GB/s):
\emph{Method~1} uses a single gp3 volume (1~disk);
\emph{Method~2} separates the WAL onto a dedicated disk (WAL: 10K~IOPS, 0.7\,GB/s; DATA: 70K~IOPS, 1.3\,GB/s);
\emph{Method~3} is the SteelDB configuration with 4~disks---WAL (10K~IOPS, 0.7\,GB/s), and three tablespaces (30K~IOPS, 0.6\,GB/s; 25K~IOPS, 0.5\,GB/s; 15K~IOPS, 0.2\,GB/s).

\vspace{-5pt} 
\section{Evaluation and Future Work}
Our eBPF analysis reveals that Method~1 loses a substantial fraction of its provisioned I/O capacity to kernel-internal stalls that are entirely invisible to standard profiling tools.
The impact on database performance is significant: as Table~\ref{tab:tpcc} shows, Method~3 (SteelDB), which eliminates these stalls, delivers the highest throughput---a 23\% improvement over Method~1---and the average write bandwidth improves by 21.7\%.
More tellingly, Method~3 reduces the maximum New--Order transaction latency by 59.4\% (1,238.2\,ms $\to$ 502.5\,ms).

Figure~\ref{fig:iostat} shows the disk I/O bandwidth over time.
Despite being provisioned with the highest single-volume IOPS (80K), Method~1 fails to reach its ceiling and delivers the lowest sustained bandwidth.
Method~3, by splitting I/O paths across multiple devices, momentarily reaches the provisioned bandwidth ceiling.
The cause becomes clear in Figure~\ref{fig:kft}, which separates I/O by issuing context.
Method~3 achieves the highest KFT writeback throughput because its three DATA disks provide three independent KFTs that drain dirty pages in parallel.
In Method~1, the single KFT is saturated; dirty pages back up, and the majority of writeback is instead driven by foreground processes through checkpoint \texttt{fsync} and other synchronous paths---an inherently less efficient mechanism that further contends with user-facing I/O.
This KFT saturation triggers the final link in the causal chain: IO-less Dirty Throttle.
Figure~\ref{fig:pause} plots the pause events over time.
Method~1, with its overwhelmed single KFT, suffers 2,513 pause episodes across the run; Method~2 reduces this to 2,094 through WAL separation, but the DATA-side single-KFT bottleneck persists.
Method~3 records only 755 pauses---a 70.0\% (2,513 $\to$ 755) reduction from Method~1---and these residual pauses are notably sparse and low in amplitude.
The substantial reduction in pauses directly explains the 59.4\% improvement in maximum latency: the tail stalls that dominated Method~1's worst-case response times are suppressed.
 
In this paper, we leverage eBPF to quantitatively identify kernel I/O bottlenecks in cloud database environments---invisible
yet dominant overheads that standard profiling tools fail to capture.
As future work, we will further investigate these cloud-specific kernel I/O bottlenecks and explore their relationships with existing kernel tuning parameters.

\bibliographystyle{ACM-Reference-Format}
\bibliography{steeldb20250317}

\end{document}